\documentclass[aps,prb,twocolumn,floatfix,superscriptaddress,longbibliography]{revtex4-2}
\usepackage{lipsum}
\usepackage{bm}
\usepackage{subcaption}
\usepackage{colordvi}
\usepackage{color}
\usepackage{comment}
\usepackage{amssymb}
\usepackage[utf8]{inputenc}
\usepackage{amsbsy, amstext, amscd, amsxtra, amsopn, amsfonts, amsthm, amsmath}
\usepackage{hyperref}
\hypersetup{colorlinks=true, linkcolor=blue, citecolor=blue, linktoc=page}
\usepackage{indentfirst}
\usepackage[final]{graphicx}
\graphicspath{ {./images/} }
\usepackage{ragged2e}
\usepackage{titlesec}
\usepackage{cancel}
\usepackage{enumitem}
\usepackage{caption}
\usepackage{breqn}
\usepackage{braket}
\usepackage{multirow}
\usepackage{siunitx}
\usepackage{soul}

\graphicspath{ {./images/} }
\usepackage[labelfont=bf, justification=Justified, format=plain]{caption}
\setcitestyle{numbers,open={[},close={]}}

\makeatletter
\let\cat@comma@active\@empty
\makeatother

\def\bn{\bar n}

\def\bOm{\bar\Omega}

\newcommand{\cO}{{\cal O}}

\newcommand{\bl}{{\overline{l}}}
\newcommand{\br}{{\overline{r}}}

\newcommand{\bS}{\bar S}

\newcommand{\bg}{{\bar g}}

\begin{document}

\title{Tunneling Dynamics and Time Delay in Electron Transport through Time-Dependent Barriers with Finite-Bandwidth Reservoirs}

\author{Shmuel Gurvitz}
\email[Corresponding author: ]{shmuel.gurvitz@weizmann.ac.il}
\affiliation{Weizmann Institute, Rehovot 76100, Israel}
\affiliation{BCAM--Basque Center for Applied Mathematics,  48009 Bilbao, Basque Country - Spain}
\author{Dmitri Sokolovski}
\email[Corresponding author: ]{dgsokol15@gmail.com}
\affiliation{Departmento de Química-Física, Universidad del País Vasco, UPV/EHU, 48940 Leioa, Spain}
\affiliation{IKERBASQUE, Basque Foundation for Science, Plaza Euskadi 5, 48009 Bilbao, Spain}
\date{\today}
\setlength{\parindent}{20pt}

\begin{abstract}
We study a model system consisting of a tunneling barrier driven by an external harmonic field and coupled to two leads with finite bandwidth. Avoiding Floquet expansions, we derive simple expressions for the time-dependent tunneling current in the adiabatic regime. Our approach relates the barrier modulation to a measurable time delay in the steady-state periodic current. It provides a physically consistent definition of the tunneling time inside the barrier by subtracting the time delay associated with the leads from the total time delay. We find that the tunneling time always vanishes for wide/high barriers. Remarkably, the time delay persists even when the barrier becomes static, i.e., in the limit where the modulation frequency vanishes. This indicates that the time delay obtained through the introduction of an external periodic perturbation actually reflects an intrinsic property of the tunneling dynamics, rather than an effect of the external drive or of a particular system. We apply our results to the analysis of tunneling times in optical experiments and find good agreement with the experimental data.
\end{abstract}

\maketitle
\section{Introduction}
\begin{figure}[h]
\includegraphics[width=6cm]{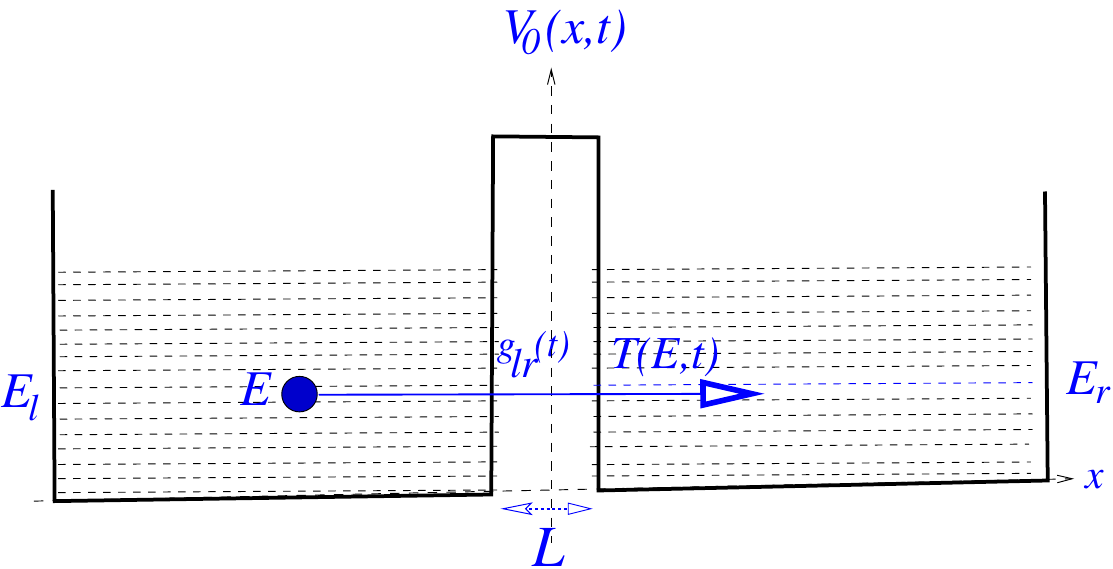}
\caption{Electron transport from the left to the right reservoir through a time-dependent barrier $V_0(x,t)$. The quantity $T(E,t)$ denotes the time-dependent single-electron transmission coefficient and corresponds to the conductance entering the Landauer formula [Eq.~\eqref{r19}].}
\label{mfig0}
\end{figure}

Electron transport through time-dependent barriers has attracted considerable attention in recent years. The analysis of time-dependent currents is commonly based on the nonequilibrium Green's function formalism \cite{jauho,croy}, often combined with Floquet theory. Although powerful, these methods are typically mathematically involved and can be computationally demanding. As a more transparent alternative, the single-electron approach (SEA) provides a conceptually simple framework that remains applicable to arbitrary time-dependent driving \cite{single,g5,g6}. In the present work, we apply the SEA to electron transport through a time-dependent tunneling barrier separating two reservoirs of finite bandwidth, as illustrated in Fig.~\ref{mfig0}.

Our results are further applied to the long-standing tunneling-time problem, which may be loosely formulated by the question: ``How long does quantum tunneling take?'' This issue has remained controversial since the pioneering work of MacColl \cite{macColl}, who concluded that the tunneling time is zero, or nearly zero, by comparing the arrival times of wave packets propagating with and without a tunneling barrier. Later, Hartman \cite{hartmann} predicted a finite tunneling time that becomes independent of the barrier width for sufficiently thick barriers, a result now commonly associated with the Hartman effect. Experimental investigations have likewise led to apparently contradictory conclusions \cite{eli}.

To date, however, no unique theoretical or experimental definition of the tunneling duration has been universally accepted, nor has such a duration been unambiguously identified with an intrinsic property of the quantum system itself \cite{REV6,eli1,phil,DSU}. Instead, tunneling times are generally inferred by applying an external perturbation or by adopting a particular measurement protocol. In this spirit, B"uttiker and Landauer (BL) proposed determining the tunneling (or traversal) time by introducing a weak harmonic modulation of the barrier height $\bar V_0(x)$ \cite{landbut,landmar}. The barrier potential then takes the form
\begin{align}
V_0(x,t)=\bar V_0(x)\left[1+\alpha \sin(\omega t)\right],
\label{harm0}
\end{align}
where $|\alpha|\ll 1$. 

In the presence of this modulation, a tunneling electron can absorb or emit energy quanta $\pm n\omega$. Using a Floquet expansion and retaining the lowest sidebands, $n=\pm1$, BL introduced a semiclassical (imaginary) tunneling time. This BL time, which is closely related to the Pollak-Miller time \cite{elim}, has subsequently been analyzed in numerous studies. In particular, it has been argued that it should not be interpreted as the ``actual'' time spent by a particle inside the barrier \cite{eli1,phil,DSU}.

Nevertheless, the idea of probing tunneling dynamics through a weak harmonic modulation of the barrier remains highly appealing. Here, we apply this idea directly to the steady-state tunneling current between the left and right reservoirs, described by the Landauer formula \cite{land} (see Fig.~\ref{mfig0}),
\begin{align}
I(t)=\frac{1}{2\pi}\int T(E,t)[f(\mu_L^{},E)-f(\mu_R^{},E)]dE\, ,
\label{r19}
\end{align}
where $f(\mu_{L,R},E)$ are the Fermi distribution functions and $\mu_{L,R}$ are the corresponding chemical potentials. Throughout this work, we set $\hbar=1$ and the electron charge $e=1$.

This formulation allows us to avoid assigning a traversal time to a particular wave packet propagating through the barrier. Instead, the tunneling dynamics is inferred from the response of the measurable electric current. An additional advantage is that the current can be measured in the leads, far from the barrier, without directly perturbing the tunneling dynamics inside the barrier region.

For a static barrier, the time-dependent transmission coefficient approaches a constant steady-state value, $T(E,t)\to T_0(E)$ for $t\to\infty$, independently of the initial condition. When the barrier is harmonically modulated at a sufficiently small frequency $\omega$ [Eq.~\eqref{harm0}], the barrier changes only weakly during the characteristic tunneling process. The steady-state transmission therefore follows the slow oscillations of the barrier, but generally with a phase shift,
\begin{align}
T(E,t)\longrightarrow
T_0(E)\left[
1+\bar\alpha\sin(\omega t-\varphi_0)
+\mathcal{O}(\bar\alpha^2)
\right].
\label{r19a}
\end{align}

The phase shift $\varphi_0$ of the induced oscillations defines a characteristic delay time $t_0=\varphi_0/\omega$. Physically, $t_0$ 
characterizes the delay between the modulation of the barrier and the resulting modulation of the transmitted current. This delay can be associated with the time interval between an electron reaches the barrier and reappears behind it. We therefore refer to $t_0$ as the ``tunneling time.'' It should be contrasted with the free-flight time $L/v$  (Fig.~\ref{mfig0}) where $v$ is the electron velocity in free space. The latter is recovered for a Markovian (wideband) reservoir.

An important feature of this definition is that the tunneling delay is extracted from the time-dependent current and therefore represents an ensemble-averaged quantity that yields only the average value of the time delay, $t_0$. This is in fully consistent with quantum mechanics which predicts only statistical averages (and correlation functions), rather than outcome of individual events.

We also emphasize the essential distinction between free propagation and tunneling through a classically forbidden region. Free propagation can be associated with the displacement of a localized wave-packet maximum. Tunneling, in contrast, proceeds through evanescent modes, for which such a simple picture of a localized peak propagating across the barrier is generally inappropriate. The transmitted wave packet results from the coherent evolution and reshaping of the wave function, which is already  spread throughout the barrier region at any time. Consequently, identifying a tunneling duration with the motion of a well-defined wave-packet peak inside the barrier would be  misleading.

Note that the tunneling delay $t_0$ is determined from oscillations of the steady-state current measured in the right lead. For a non-Markovian reservoir of finite bandwidth, propagation in the lead itself carries a memory-induced delay. The measured $t_0$ can therefore contain contributions both from the tunneling barrier and from the finite-bandwidth reservoir. Isolating the intrinsic barrier contribution would appear to require subtracting the lead-induced delay, thereby reintroducing one of the central ambiguities of the tunneling-time problem \cite{eli,eli1}. In the following, we show that our procedure  provides an unambiguous procedure for separating these two contributions and thereby extracting the tunneling delay associated specifically with the barrier.

\section{Tunneling Hamiltonian}

For analysis of the tunneling transport, we use the tunneling Hamiltonian (TH) approach, which has been widely used in mesoscopic transport \cite{datta} and in solid-state physics, in general. Thus, we describe an entire system, shown in Fig.~\ref{mfig0} by the following Hamiltonian 
\begin{align}
H(t)=H_L^{}+H_R^{}+\sum_{l,r}g_{lr}^{}(t)\Big[|l\rangle\langle r|+|r\rangle\langle l|\Big]   
\label{apm1}
\end{align}
where $H_{L(R)}=\sum_{l(r)}E_{l(r)}^{}\ket{l(r)}\bra{l(r)}$ is the Hamiltonian of the isolated left (right) reservoir and $g_{lr}(t)$ is the time-dependent tunneling coupling (energy) between the levels $E_{l,r}$ of left/right lead. It describes electron tunneling through the barrier and is responsible for the current flow. In the absence of a magnetic field, the tunneling coupling $g_{lr}(t)$ may be chosen real. 

To find the time-dependent current, we use the SEA by solving directly the time-dependent Schr\"odinger equation for a single electron,  
\begin{align}
i\frac{\partial}{\partial t}\ket{\Psi(t)}=H(t)\ket{\Psi(t)}
\label{se}  
\end{align}
to derive analytical expressions for the time-dependent transmission coefficient (conductance) $T(E,t)$ and the tunneling time delay $t_0$ in the adiabatic limit. Note that the standard approach \cite{jauho}, based on the Floquet expansion \cite{croy}, is not well suited to the low-frequency regime of harmonic driving. In this regime, one must sum over a large number of modulation quanta, $\pm n\omega$, which considerably complicates the analysis of the transition to the adiabatic limit. This transition is of central interest in the present work.

Within the SEA, the Floquet expansion can be avoided altogether. Instead, one solves the time-dependent single-particle Schr\"odinger equation,  which describes the dynamics of an electron occupying the system shown in Fig.~\ref{mfig0}. 
The corresponding wave function can be written as  
\begin{align}
|\Psi^{(\alpha)}_{}(t)\rangle=\sum_l b_{l}^{(\alpha)}(t)|l\rangle+\sum_r b_r^{(\alpha)}(t)|r\rangle
\label{r0}
\end{align}
where $\alpha=\{\bl,\br\}$ specifies the initially occupied state.

Substituting Eq.~\eqref{r0} into the Schr\"odinger equation yields a set of coupled first-order differential equations for the amplitudes $b_{l,r}^{(\alpha)}(t)$. For an electron initially occupying the state $E_{\bar l}$ in the left reservoir, these equations can be written in the integral form as  
\small
\begin{subequations}
\label{cr2}
\begin{align}
&b_l^{(\bl)}(t)=e^{-iE_l^{}t}\delta_{l\bl}-i\int\limits_0^t\sum_r g_{lr}^{}(t')b_r^{(\bl)}(t')e^{-iE_l^{}(t-t')}\,dt'
\label{cr2a}\\
&b_r^{(\bl)}(t)=-i\int\limits_0^t\sum_l
g_{rl}^{}(t')b_l^{(\bl)}(t')e^{-i E_r^{}(t-t')}\, dt'
\label{cr2b}
\end{align}
\end{subequations}
\normalsize
The amplitudes $b_{l(r)}^{(\bar l)}(t)$ completely determine the dynamics of a non-interacting many-electron system \cite{g5,g6}. In particular, the many-electron current can be expressed in terms of single-electron current evaluated for initially empty reservoirs. The current flowing into the right reservoir from an electron initially occupying the state $E\equiv E_{\bar l}$ in the left reservoir is $i_{R}^{}(E,t)=\sum_r (d/dt)|b_r^{}(E,t)|^2$, where $b_{l(r)}^{}(E,t)\equiv b_{l(r)}^{(\bl)}(t)$. 

In the continuum limit, the sums over reservoir states are replaced by energy integrals weighted by the density of states, e.g., $\sum_r\to\int\rho(E_r)dE_r$. If all states within an energy interval $\Delta E$ around energy $E$ in the left reservoir are occupied, the corresponding current in the right reservoir is obtained by summing the contributions of all occupied states, $I_{R}^{}(t)=T(E,t)\Delta E$, where
\small 
\begin{align}
T(E,t)=\rho(E)i_R(E,t)=\rho(E)\int\frac{d}{dt}|b_r^{}(E,t)|^2\rho(E_r)dE_r^{}
\label{r22}
\end{align}
is the time-dependent transmission coefficient (or conductance) \cite{g6}. Using Eq.\eqref{cr2} one obtains 
\small
\begin{align}
\frac{d}{dt}|b_r^{}(E,t)|^2=2{\rm Re}\int\limits_0^t\sum_{l,l'}g_{lr}^{}(t)g^{}_{l'r}(t')&b_l^{}(E,t)b_{l'}^{*}(E,t')
\nonumber\\
&\times 
e^{iE_r^{}(t-t')}dt'
\label{r22p}
\end{align}
\normalsize
where $b_{l(l')}(E,t)\equiv b_{l(l')}^{(\bar l)}(t)$.  In the limit of a static barrier and Markovian reservoirs (infinite bandwidth), $T(E)$ reduces to the familiar transmission probability entering the Landauer formula [Eq.\eqref{r19}], which determines the linear response (conductance) of the barrier.

\section{Bardeen formula and applicability of the TH}

We emphasize that the above derivation is formulated entirely within the TH framework [Eq.~\eqref{apm1}], in which the spatial dynamics inside the barrier does not appear explicitly. This does not imply, however, that the under-barrier dynamics is neglected. Rather, information about the barrier and the evanescent propagation through it is encoded in the tunneling matrix element $g_{lr}(t)$. Indeed, starting from the Schr"odinger equation in coordinate space, the tunneling matrix element can be expressed through the Bardeen formula \cite{bard,bard1},
\begin{align}
g_{lr}(t)=\frac{1}{2m}
\left[
\chi_l'(x,t)\chi_r(x,t)
-\chi_l(x,t)\chi_r'(x,t)
\right]{x\to0},
\label{12}
\end{align}
where $\chi{l,r}(x,t)$ are the wave functions associated with the left and right reservoirs, evaluated at a surface inside the barrier (chosen here at its midpoint, $x=0$), and $\chi'(x,t)\equiv\partial_x\chi(x,t)$.

Since these wave functions decay exponentially inside the classically forbidden region, the tunneling coupling $g$ is strongly suppressed for a sufficiently high or wide barrier and depends explicitly on the instantaneous barrier profile. This can be seen by evaluating Eq.~\eqref{12} semiclassically. Using the semi-classical wave functions for $\chi_{l,r}(x,t)$ in the Bardeen formula, together with Eqs.~\eqref{cr2} and \eqref{r22} for Markovian (wideband) reservoirs, one obtains the transmission coefficient \cite{wil}
\begin{align}
T(E,t)=e^{-2\int\limits_{-x_0^{}}^{x_0^{}}\sqrt{2m[V_0^{}(x',t)-E]}dx'}\, ,
\label{sl}
\end{align}
where $\pm x_0$ are the turning points on the barrier ($x_0=L/2$ in Fig.~\ref{mfig0}). 

Equation~\eqref{sl} makes explicit the connection between the TH description and evanescent propagation inside the barrier. The transmission probability is exponentially suppressed as the barrier height or width increases. Consequently, $T\ll1$ provides a natural small parameter for a controlled perturbative derivation of the tunneling Hamiltonian. More formally, both the TH and the Bardeen formula can be obtained from the coordinate-space Schr"odinger equation using the modified perturbative expansion known as the ``two-potential approach'' \cite{g1,g2}, in which the expansion is organized in powers of the barrier penetration factor $T\sim |g|^2$ \cite{gur3,gur4}. The leading transmission probability is of order $g^2\sim T$, whereas the first neglected corrections are of order $g^4\sim T^2$. Thus, the relative accuracy of the TH description improves exponentially as the barrier becomes higher or wider.

Since the TH is accurate up to the $g^2$ terms, it is not necessary to keep the higher power terms in the amplitudes $b_{l,r}^{}(E,t)$, given by Eqs. \eqref{cr2} and \eqref{r22}. It allows us to neglect the second term in  Eq. \eqref{cr2a} by replacing $b_l^{(\bl)}(t)\simeq e^{-iEt}\delta_{l\bl}$
in Eq.~\eqref{cr2b}. Substituting the resulting amplitudes into Eq.~\eqref{r22} yields
\small
\begin{align}
T(E,t)=2\rho(E)\,{\rm Re}\int\limits_0^t d\tau\int\limits_{-\infty}^\infty S_r^{}(E,t,\tau)e^{i(E_r^{}-E)\tau}dE_r
\label{r20}
\end{align}
\normalsize
where
\begin{align}
S_r(E,t,\tau)=g_r^{}(E,t)g^{}_r(E,t-\tau)\rho(E_r)
\label{r20a}
\end{align}
with $g_r(E,t)\equiv g_{\bar l r}(t)$. The quantity $S_r(E,t,\tau)$ plays the role of a time-dependent spectral function of the right reservoir \cite{datta}. Correspondingly,
\begin{align}
C(E,t,\tau)= \int\limits_{-\infty}^\infty S_r^{}(E,t,\tau)e^{i(E_r-E)\tau}dE_r 
\label{r20b}
\end{align}
is the reservoir correlation function in the presence of the time-dependent barrier. Its dependence on $g_r(E,t-\tau)$ explicitly shows that the current at time $t$ can retain information about the barrier coupling at earlier times, thereby incorporating the reservoir memory.

Equation~\eqref{r20} is the central analytical result of this section. It expresses the time-dependent transmission coefficient directly in terms of the reservoir spectral function and remains applicable to arbitrary temporal modulation of the barrier within the accuracy of the TH approximation. Corrections of order $g^4\sim T^2$ and higher lie beyond this accuracy \cite{datta,gur3}.

In our treatment, electrons in the leads are assumed to be noninteracting, or, equivalently, electron-electron interactions are included only at the self-consistent mean-field level, as is commonly done in theoretical studies of time-dependent electron transport \cite{jauho,croy}. Interaction effects may become important when the system strongly departs from the Markovian limit, for example by effective (``fictitious'') quantum wells in reservoirs \cite{g6}, in which case electron-electron interaction may no longer be negligible (like in Kondo effect). A systematic treatment of such corrections lies beyond the scope of the present analysis.

The decoherence effects due to random fluctuations of the barrier potential can be incorporated within the SEA framework as in Refs.~\cite{g5,g6}. However, such fluctuations may therefore be regarded as part of the time-dependent barrier potential. Consequently, they do not affect (on average) the time-dependent current oscillations considered here, and therefore are not considered in the present analysis. 

\section{Markovian leads and tunneling time}

We now use Eq.~\eqref{r20} to evaluate the delay time $t_0$. We first consider Markovian reservoirs, corresponding to the wideband limit. In this limit, the spectral function $S_r(E,t,\tau)$ of Eq.~\eqref{r20a} can be taken to be independent of the reservoir energy $E_r$. The energy integral in Eq.~\eqref{r20b} then gives a delta-correlated reservoir response, $C(E,t,\tau) \propto \delta (\tau)$,
so that the reservoir has no memory.

As a consequence, Eq.~\eqref{r20} becomes local in time: The transmission coefficient at time $t$ is determined by the instantaneous value of the tunneling coupling. For the slowly modulated barrier of Eq.~\eqref{harm0}, the steady-state transmission therefore follows the barrier modulation without a phase lag. Comparing with Eq.~\eqref{r19a}, we obtain $\varphi_0=0, \quad t_0=\varphi_0/\omega=0$. Thus, within the accuracy of the TH approximation, a Markovian reservoir produces neither a reservoir-memory delay nor a finite phase delay associated with the tunneling barrier. In terms of the operational definition introduced above, the tunneling time therefore vanishes, $t_0=0+\mathcal{O}(T^2)$.

It is noteworthy that this result is consistent with the conclusion reached by MacColl \cite{macColl}, although the two approaches are conceptually very different. MacColl inferred a vanishing, or nearly vanishing, tunneling delay by comparing wave-packet arrival times in the presence and absence of a barrier. Here, in contrast, the delay is extracted from the phase of the steady-state current induced by a weak periodic modulation of the barrier. Within the TH framework, the result does not depend on the detailed shape of an incident wave packet and, in the Markovian limit, is independent of the particular barrier profile to leading order in the transmission probability.

This last point is especially important. The leading TH contribution gives no phase delay, while corrections arise only beyond the order retained in the tunneling Hamiltonian, namely, at order $T^2$. Since $T$ decreases exponentially with increasing barrier height or width, any nonzero correction to the phase delay is correspondingly suppressed for an increasingly opaque barrier. In this sense, the present result is consistent with the saturation underlying the Hartman effect \cite{hartmann}; within the leading TH approximation, this limiting delay is zero.

A vanishing tunneling delay may appear counterintuitive, particularly when interpreted in terms of a classical traversal picture, and its relation to causality and relativistic propagation has been discussed extensively in the literature \cite{eli}. We do not address these broader interpretational questions here. Instead, our purpose is to establish the operational result relevant to the present formulation: For Markovian leads, the phase delay of the tunneling current and tunneling time vanish within the accuracy of the TH approach. This provides a particularly useful reference point for the analysis below, where finite reservoir bandwidths introduce memory effects and hence a nonzero contribution to the measured delay. We shall then examine how this reservoir-induced contribution can be distinguished from that associated with tunneling through coupled barriers.

\section{Non-Markovian leads and Kronig–Penne model}

\begin{figure}[h]
\includegraphics[width=8cm]{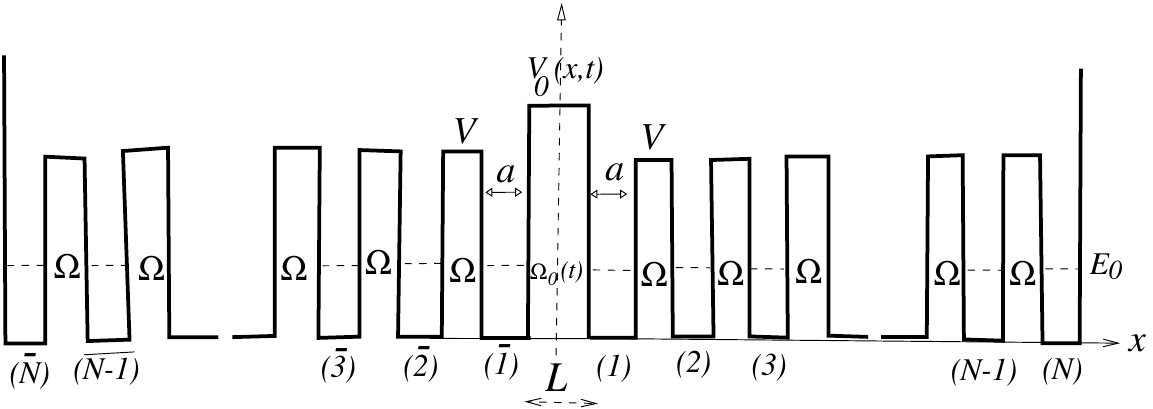}
\caption{Two one-dimensional periodic lattices representing the left and right reservoirs, separated by the time-dependent barrier $V_0(x,t)$. Each lattice consists of identical quantum wells of width $a$. Nearest-neighbor wells within each lead are coupled by $\Omega$, while the two boundary wells adjacent to the central barrier are coupled by the time-dependent amplitude $\Omega_0(t)$.}
\label{mfig2}
\end{figure}

We now turn to the non-Markovian contribution to the time-dependent tunneling current. Equation~\eqref{r20} provides a convenient framework for investigating how a finite reservoir bandwidth, and the resulting memory effects, modifies the time-dependent conductance. To this end, we model each reservoir as a one-dimensional Kronig-Penney lattice consisting of $N$ identical quantum wells, as illustrated in Fig.~\ref{mfig2}. This model captures the finite-band structure of the leads while remaining analytically tractable.

Restricting the dynamics within each lead to nearest-neighbor hopping, the corresponding tunneling Hamiltonian takes the form
\begin{align}
H(t)=H_L+H_R+\Omega_0(t) \big(|1\rangle\langle \overline{1}|+|\overline{1}\rangle\langle 1|\big),
\label{2a2}
\end{align}
[cf. Eq.\eqref{apm1}], where
\begin{align}
H_L=E_0\sum_{n=1}^N|\overline{n}\rangle\langle \overline{n}|+ \Omega \sum_{n =1}^{N-1}\big(|\overline{n}\rangle\langle \overline{n+1}|+H.c.\big),
\label{2a3}
\end{align}
describes the left lead. Here, $E_0$ is the energy of an isolated well and $\Omega$ is the nearest-neighbor hopping amplitude. The states $|\overline n\rangle$ correspond to the quantum wells located to the left of the time-dependent barrier $V_0(x,t)$, as shown in Fig.~\ref{mfig2}. The Hamiltonian $H_R$ of the right lead is defined analogously by replacing $|\overline n\rangle$ with $|n\rangle$.

The last term in Eq.~\eqref{2a2} couples the boundary wells $|\overline1\rangle$ and $|1\rangle$ of the two leads across the time-dependent barrier $V_0(x,t)$ of Eq.~\eqref{harm0}. Its coupling amplitude $\Omega_0(t)$ therefore inherits the temporal modulation of the barrier, whereas $\Omega$ characterizes tunneling between neighboring wells within each reservoir.

Both hopping amplitudes, $\Omega$ and $\Omega_0(t)$, can be evaluated using the Bardeen formula, [Eq.~\eqref{12}]. In this case, the reservoir wave functions $\chi_{l,r}$ entering that formula are replaced by the localized wave functions of the quantum wells adjacent to the corresponding barrier. Thus, $\Omega$ is determined by the overlap of neighboring well states within a lead, while $\Omega_0(t)$ is determined by the overlap of the two boundary-well states across the central, time-dependent barrier.

Diagonalization of $H_L$ yields the eigenenergies and eigenstates of the left lead [Fig.~\ref{mfig0}, Eq.\eqref{apm1}]
\begin{subequations}
\label{2a5}
\begin{align}
&E_l=E_0+2\Omega\cos \big[\pi l/(N+1)\big]\ \text{for}\ l=1,\ldots,N
\label{2a5a}\\
&\ket{l} =\sqrt{2/(N+1)}\sum_{n=1}^N\sin [\pi l\, n/(N+1)\big]\ket{\bn}.
\label{2a5b}
\end{align}
\end{subequations}
The corresponding expressions for the right lead are obtained by replacing $\ket{\overline n}\rightarrow\ket n$ and $l\rightarrow r$. 

In the continuum limit, the density of states in the left lead is
\begin{align}
\rho(E_l)=-\left(dE_l^{}/dl \right )^{-1}=(N+1)/[2\pi\Omega\, \eta(E_l)]
\label{a9}
\end{align}

where 
\begin{align}
\eta(E_l)= \sqrt{1-\Big(\frac{E_l-E_0}{2\Omega}\Big)^2}
\label{eta}
\end{align}for $(E_l-E_0)^2\le 4\Omega^2$, and $\eta(E_l)=0$ otherwise. An identical expression holds for the right lead after replacing $E_l$ by $E_r$.

The tunneling amplitudes $g_{lr}(t)$ entering Eq.~\eqref{apm1} can now be expressed in terms of the inter-lead coupling $\Omega_0(t)$. Expanding the boundary states $\ket{\overline1}$ and $\ket{1}$ in the eigenstate bases of the left and right leads, respectively, yields
\begin{align}
g_{lr}^{}(t)=2\Omega_0(t)\eta(E_l^{})\eta(E_r^{})/(N+1).
\label{2a9}
\end{align}
Substituting Eq.~\eqref{2a9} into Eqs.~\eqref{r20} and \eqref{r20a}, we obtain 
\begin{align}
2\rho(E)S_r(E,t,\tau)=2\,\eta(E)\frac{\Omega_0(t)\Omega_0(t-\tau)}{\pi^2\Omega^2}\eta(E_r^{})\, ,
\label{spd}
\end{align}
The energy dependence is therefore entirely determined by the factor $\eta(E)$ [Eq.~\eqref{eta}]. As a result, the Kronig-Penney model generates the well-known semicircular spectral density of a finite support $4\Omega$ \cite{yijin}.

Substituting Eq.~\eqref{spd} into Eq.~\eqref{r20} and performing the integration over $E_r$, we arrive at
\small
\begin{align}
T(E,t)=\frac{2\eta(E)}{\pi}\int\limits_0^t \frac{\Omega_0(t)\Omega_0(t-\tau)}{\Omega^2}\frac{J_1(2\Omega\tau)}{\tau}
\cos(E\tau)d\tau
\label{rr20}
\end{align}
\normalsize
where $J_1(2\Omega\tau)\cos (E\tau)/\tau$ represents the bath correlator, $C(E,\tau)$ [Eq.\eqref{r20b}], in terms of the Bessel function of the first kind.

We now evaluate the time-dependent conductance $T(E,t)$ for a barrier undergoing weak harmonic modulation [Eq.~\eqref{harm0}]. Using the Bardeen formula, Eqs.~\eqref{12} and \eqref{sl} for the tunneling coupling, we can write 
\begin{align}
\Omega_0^{}(t)=\bOm_0[1+\bar\alpha\sin (\omega t)+{\cO(\bar\alpha^2)]}\, ,
\label{harm}
\end{align}
where $|\bar\alpha|\ll1$ and $\bar\Omega_0$ is the static component of the tunneling coupling.

Substituting Eq.~\eqref{harm} into Eq.~\eqref{rr20} and retaining only terms linear in $\bar\alpha$, we consider the steady-state regime. In this limit, the upper integration limit may be extended to infinity, since the memory kernel $J_1(2\Omega\tau)/\tau$ decays on a timescale of order $\Omega^{-1}$. Evaluating the integral in the adiabatic limit, $\omega/(2\Omega)\ll1$, yields
\begin{align}
&T(E,t)=\frac{2\bOm_0^2}{\pi\Omega^2}\eta^2(E)\Big[1+2\bar\alpha\sin[\omega (t -t_0)]\Big],
\label{nm2}\\
&\text{where}\nonumber\\
&t_0=\frac{1}{\omega}\arcsin\Big(\frac{\omega}{2\Omega\eta(E)}\Big){\xrightarrow {\omega\to 0}}\frac{1}{2\Omega\sqrt{1-E^2/(2\Omega)^2}}.
\label{nm3}
\end{align}
Here and below, we set $E_0=0$. 

Equations~\eqref{nm2} and \eqref{nm3} show that the conductance oscillates with the same frequency as the barrier but lags behind it by a phase shift $\omega\, t_0$, indicating that the delay is inversely proportional to the reservoir bandwidth, or equivalently to the nearest-neighbor hopping amplitude $\Omega$ of the lattice model (cf.~ Ref.\cite{landmar}). 

A similar result is obtained for spectral densities with infinite support. As an example, consider the Lorentzian spectral function frequently employed in transport theory \cite{Lorentz},
\begin{align}
\bS_r(E,t,\tau)=\frac{\bg^2(E)}{\pi}\frac{\Lambda}{\Lambda_{}^2+E_r^2}[1+\bar\alpha{\cal F}(t,\tau)]  
\label{lorentz}
\end{align}
where ${\cal F}(t,\tau)=\sin\omega t+\sin\omega (t-\tau)$ and $|\bar\alpha|\ll1$. The parameter $\Lambda$ sets the effective bandwidth of the reservoir. As before, terms of order $\bar\alpha^2$ are neglected. For $\Lambda=2^{3/2}\Omega$, the Lorentzian spectral density has the same curvature at the band center as the semicircular density generated by the Kronig-Penney model.

Substituting Eq.~\eqref{lorentz} into Eqs.~\eqref{r20} and \eqref{r20a} and performing the integrations analytically, we obtain
\small
\begin{align}
t_0=\frac{1}{\omega}\arcsin\Big(\frac{\omega}{2\Lambda}\frac{\Lambda^2-E^2+\omega^2}{\Lambda^2+E^2+\omega^2}\Big)
{\xrightarrow {\omega\to 0}}\frac{1}{2\Lambda}\frac{1-\big(\frac{E}{\Lambda}\big)^2_{}}{1+\big(\frac{E}{\Lambda}\big)^2_{}}.
\label{nm4}
\end{align}
\normalsize 
Thus, despite its infinite support, the Lorentzian spectral density produces essentially the same delay near the band center ($|E|/\Lambda\ll1$) as the finite-bandwidth Kronig-Penney model, with only a weak dependence on energy.

\section{Tunneling through a multi-barrier system}

The finite tunneling delay $t_0$ obtained for a barrier coupled to non-Markovian leads, [Eqs.~\eqref{nm3} and \eqref{nm4}] raises an important question concerning its physical origin. In particular, does this finite delay result from a modification of the under-barrier dynamics by the non-Markovian reservoirs, or does it originate from electron propagation within the finite-bandwidth leads themselves?

\begin{figure}[h]
\includegraphics[width=8.5cm]{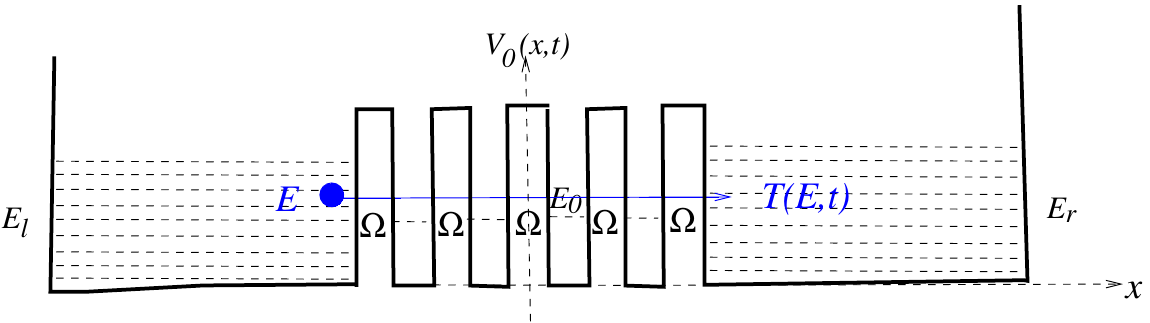}
\caption{Periodic multi-barrier system coupled to two Markovian leads. One of the barriers is harmonically modulated by an external periodic field.}
\label{mfig3}
\end{figure}

To clarify this point, we consider a periodic multi-barrier system coupled to Markovian leads, as shown in Fig.~\ref{mfig3}.

The system can again be described within the TH framework of Eq.~\eqref{apm1}. In this case, however, the term describing tunneling through a single barrier is replaced by a finite chain of coupled wells of the form appearing in Eq.~\eqref{2a3}. The resulting structure is essentially a finite Kronig-Penney system consisting of $N$ wells separated by tunneling barriers. For finite $N$, the corresponding energy band is represented by the discrete set of levels given in Eq.~\eqref{2a5}. For sufficiently large $N$, however, the level spacing becomes small and the continuum limit, $N\to\infty$, can be used to evaluate the time-dependent transmission coefficient $T(E,t)$ and, consequently, the tunneling current of Eq.~\eqref{r19}.

The geometry in Fig.~\ref{mfig3} is closely related to that of Fig.~\ref{mfig2}, with the important difference that the finite periodic structure is now connected to Markovian leads. The phase delay of the time-dependent tunneling current can therefore be evaluated in essentially the same way as for the non-Markovian reservoir considered above. In the low-frequency limit and near the center of the band, we obtain $t_0\simeq 1/|2\Omega|$ [cf. Eq.~\eqref{nm3}]. Thus, in contrast to tunneling through a single barrier coupled to Markovian leads, for which the phase delay vanishes within the TH approximation, the multi-barrier system exhibits a finite delay even though the external leads themselves are Markovian.

This difference provides insight into the physical origin of the delay. For a single sufficiently opaque barrier, transport through the classically forbidden region is mediated exclusively by evanescent modes, as discussed in the Introduction. Such modes do not correspond to the conventional propagation of a localized wave-packet maximum. Rather, transmission results from the coherent reshaping of the wave function across the barrier. This process can produce an anomalously small phase delay, reflecting the nonlocal correlations associated with evanescent-wave propagation \cite{chiao2,chiao,chiao1,sok,xin}. Within the TH approximation considered here, the corresponding delay tends to zero as the barrier becomes increasingly opaque, i.e., as its width or height is increased.

The situation is qualitatively different for a sequence of coupled barriers. In the multi-barrier structure of Fig.~\ref{mfig3}, the electron does not propagate exclusively through classically forbidden regions. Between successive barriers, its propagates through classically allowed region. This dynamics is incorporated implicitly in the microscopic derivation of the Kronig-Penney tunneling Hamiltonian \cite{gur3}. Consequently, the system possesses an intrinsic dynamical time scale of order $\sim 1/|2\Omega|$. This explains the finite tunneling time, in contrast to the vanishing $t_0$ obtained for an isolated barrier coupled to Markovian reservoirs.

This observation also provides a natural interpretation of the finite delay found for a single barrier connected to non-Markovian leads (Fig.~\ref{mfig2}). We propose that the nonzero $t_0$ in that case does not originate from a modification of the evanescent dynamics inside the barrier itself. Rather, it reflects electron dynamics in the finite-bandwidth reservoirs, in close analogy with the propagation through the intermediate wells of the multibarrier system in Fig.~\ref{mfig3}. This interpretation is supported directly by Eqs.~\eqref{nm3} and \eqref{nm4}: The resulting delay is determined by the reservoir bandwidth (or, equivalently, by the hopping amplitude $\Omega$) and does not depend explicitly on the parameters of the central tunneling barrier.

The comparison between Figs.~\ref{mfig2} and \ref{mfig3} therefore suggests a useful separation of timescales. The phase delay associated with an opaque individual barrier vanishes to the accuracy of the TH approximation, whereas a finite delay can arise from propagation in regions supporting non-evanescent dynamics, such as the wells between successive barriers or finite-bandwidth reservoirs. In this sense, the non-Markovian contribution to $t_0$ can be interpreted as a propagation or memory delay of the leads rather than as an intrinsic traversal time of the central barrier.

These predictions could, in principle, be tested in systems in which the bandwidth of the reservoirs can be varied experimentally. Changing the bandwidth would directly modify the reservoir-memory contribution predicted by Eqs.~\eqref{nm3} and \eqref{nm4}, while leaving the central barrier unchanged. Conversely, the prediction of a vanishing phase delay for a single barrier coupled to Markovian reservoirs could be examined by systematically increasing the barrier width or height and determining whether the measured delay approaches zero in the opaque-barrier limit. Such a measurement would provide a direct test of the tunneling-time interpretation developed here and of its relation to the Hartman effect \cite{hartmann}.

\section{Generalization to different systems}
Perhaps the most significant result of the present work is that the delay time given by Eqs.~\eqref{nm3} and \eqref{nm4}, and hence the tunneling time inferred from it, remains finite when $\omega\to0$. At first sight, this result may appear surprising, since in this limit the barrier is a static one.

The apparent paradox arises because the external modulation is essential for revealing the time delay. Both the theoretical determination of the delay and its experimental measurement rely on the response of the tunneling current to a weak periodic perturbation. However, the corresponding delay can remain finite even though the phase shift itself vanishes linearly with frequency, $\varphi_0(\omega)\propto\omega$. Thus, the finite value of $t_0$ in the limit $\omega\to0$ is not generated by the external drive. Rather, the periodic modulation provides a clock that reveals a characteristic dynamical timescale already present in the system. This observation suggests that the results of Eqs.~\eqref{nm3} and \eqref{nm4} may have a broader applicability and could be useful for analyzing tunneling dynamics in other systems.

 As an illustration, we compare our result with the single-photon tunneling-time experiment of Steinberg, Kwiat, and Chiao \cite{chiao}, in which a two-photon interferometer was used to measure the propagation delay through a one-dimensional barrier about $d=1.1\, \mu$m thick.  The barrier was formed by a dielectric-mirror multilayer  photonic band-gap structure. One photon traversed the dielectric multilayer, while the second photon provided a reference. The transmitted photon was found to emerge $\Delta t=t_0-d/c=-1.47\pm 0.21$ fs earlier than a photon propagating the same geometrical distance $d$ in vacuum. Since $d/c\simeq 3.67$ fs, this corresponds to an experimental transit time $t_0^{\rm exp}\simeq 2.20\pm$ 0.21~fs. 

Within its stop band, propagating electromagnetic modes are suppressed and transmission occurs through evanescent modes. The multilayer structure can therefore be viewed as the optical realization of a finite Kronig–Penney lattice (Fig.~\ref{mfig3}), representing a finite sequence of coupled layers connected on both sides to radiative continua that serve as Markovian reservoirs, shown in Fig.~\ref{mfig3}. It gives $t_0\simeq 1/|2\Omega|$ [Eq.~\eqref{2a5}], where $4|\Omega|$ is the full energy support in the Kronig-Penney model. We therefore estimate it as the effective optical bandwidth from the experimentally relevant spectral interval (band gap), approximately $\lambda_1=600\ {\rm nm}$ to $\lambda_ 2=800\ {\rm nm}$, reported in Ref. \cite{chiao}. 

Converting this wavelength interval into an energy interval gives for the full support $4|\Omega|\simeq hc (1/\lambda_1-1/\lambda_2)$ and hence the tunneling time $t_0=1/|2\Omega|\simeq 2.55$ fs is reasonably close to the experimentally inferred 2.20±0.21 fs. Considering the simplified mapping of the dielectric multilayer onto a nearest-neighbor tight-binding model, the proximity of these two values is encouraging. In particular, an effective (finite range) support $|4\Omega|$ in the Kronig-Penney model should be slightly larger than an effective bandwidth, extracted  directly from the experimental transmission spectrum. It could make an agreement with $t_0^{\rm exp}$ even more pronounced.   

\section{Summary and Conclusions}
In summary, we have investigated electron transport through a harmonically modulated tunneling barrier coupled to reservoirs of finite bandwidth. Using the TH approach, we have shown that the steady-state current follows the periodic modulation of the barrier but, in general, acquires a phase shift. This phase shift defines a characteristic delay time, $t_0=\varphi_0/\omega$, which provides an operational measure of the characteristic time scales associated with tunneling through the barrier and propagation in the leads.

In the Markovian (wide-band) limit, the reservoir correlation function becomes local in time and the corresponding phase delay vanishes within the accuracy of the TH approximation. In particular, for a sufficiently high or wide single barrier, the tunneling delay approaches zero as the transmission probability becomes exponentially small. In contrast, a finite delay arises for non-Markovian reservoirs of finite bandwidth and also for multi-barrier structures connected to Markovian leads. Our analysis shows that these finite delays originate from dynamical propagation and memory effects outside the individual classically forbidden regions. This interpretation provides a natural separation between the vanishing delay associated with an opaque single barrier and the finite contribution generated by propagation in finite-bandwidth leads or in the quantum wells separating successive barriers.

An important result of our analysis is that the characteristic delay time remains finite in the limit of $\omega\to 0$, even though the phase shift itself vanishes as $\varphi_0 \propto\omega$. The weak periodic modulation therefore does not generate the characteristic timescale; rather, it acts as a probe that makes this timescale observable through the phase of the tunneling current. Consequently, the time scales obtained from our analysis characterize the underlying static tunneling system and may also be relevant to experimental protocols in which the delay is determined by other means. As an illustration, we applied our result to photon tunneling through a multilayer dielectric structure and obtained a characteristic transit time of the same femtosecond scale as that observed experimentally.

Finally, the present analysis has been formulated explicitly for one-dimensional systems. This restriction is not fundamental: Both the tunneling-Hamiltonian approach and the Bardeen formula can be generalized to multidimensional geometries \cite{fn1,bard2,gur11}. Moreover, tunneling through a multidimensional barrier is known to be dominated semiclassically by an optimal escape trajectory, with the transmitted wave packet remaining relatively narrowly distributed around this trajectory \cite{gur11}. This suggests that the physical picture developed here---in particular, the distinction between the delay associated with evanescent penetration through an individual barrier and that arising from propagation or memory outside the barrier---may extend to multidimensional tunneling as well. A detailed investigation of this generalization is left for future work.

\section*{Acknowledgements}

One of us (S.G.) like to thank Prof. Eli Pollak for helpful discussions. S.G. is grateful for financial support and hospitality offered by the Basque Center for Applied Mathematics during his visit to Bilbao in the autumn of 2023. D.S. acknowledges financial support by the Grant PID2021-126273NB-I00 funded by MICINN/AEI/10.13039/501100011033 and by {ERDF A way of making Europe}, as well as by the Basque Government Grant No. IT1470-22.


\begin{references}
\bibitem{fn1} In the 3D case it represents an integral over the surface through the tunneling barrier, separating the leads \cite{bard, gur11,bard2}.
\end{references}
\end{document}